\documentclass{webofc}
\usepackage[varg]{txfonts}   
\newcommand{\vect}[1]{{\mathbf {#1}}}
\begin{document}
 \title{Toward the measurement of the hyperfine structure of
 muonic hydrogen in the FAMU experiment.}
%
%
\subtitle{Multi-pass cavity optimization for experiments with
pulsed sources.}

\author{\firstname{Dimitar} \lastname{Bakalov}\inst{1}\fnsep\thanks{\email{dbakalov@inrne.bas.bg}} \and
        \firstname{Mihail} \lastname{Stoilov}\inst{1}\fnsep
 }

\institute{Institute for Nuclear Research and Nuclear Energy,
 Bulgarian Academy of Sciences,
 Tsarigradsko ch. 72, Sofia 1784, Bulgaria
          }

\abstract{%
 We consider a simplified model of the optical multi-pass cavity
 that is being currently developed by the FAMU collaboration
 for the measurement of the
 hyperfine splitting in the ground state of muonic hydrogen
 and of the Zemach radius of the proton.
 The model is focused on the time distribution of the
 events of laser-stimulated hyperfine transitions in the muonuc
 atom and may be helpful in the
 preliminary design of the FAMU experimental set-up
 and, more generally, in the optimization of multi-pass optical
 cavities for experiments with pulsed lasers.
}
\maketitle
\section{Introduction}
\label{intro}

 The laser
 excitation of the ortho $S=1$ hyperfine sub-level of the ground state of
 the muonic hydrogen atom from the para $S=0$ sub-level (where
 $\vect{S}=\vect{S}_p+\vect{S}_{\mu}$ is the total spin of the
 muonic hydrogen atom $\mu^-p$) is an ambitious project of the
 FAMU collaboration \cite{famu-coll}:
 the measurement of the resonance transition frequency will provide
 top-accuracy data on the Zemach radius of
 the proton\cite{famu-exp}. This is a very weak M1 magnetic dipole transition
 with probability of only
 $2\times10^{-5} (E/{\rm J}) (S/{\rm m}^2)^{-1}({\rm T}/{}^{\circ}\!K)^{-1/2}$
 for laser pulse energy $E$, laser beam cross section $S$ and
 target temperature $T$ \cite{NIMB12}, for which
 an optical multi-pass cavities needs to be used to enhance the probability
 of laser-stimulated transitions to a reasonable level.

 The measurement of the hyperfine splitting in the ground state of
 muonic hydrogen consists in the detailed study of the chain of
 processes occurring when muonic hydrogen atoms in a mixture of
 hydrogen and oxygen interact with laser radiation tuned at a
 frequency around the hyperfine transition resonance frequency
 \cite{NIMB12}. The observable quantity used as
 signature is the time distribution of the events of muon transfer
 from hydrogen to oxygen \cite{famu-mutran} recognized by the
 emission of characteristic X-rays : the maximal deviation
 of the latter from the background time distribution in absence of
 laser indicates that the laser source is tuned at the resonance
 frequency.

 The FAMU experiment will use the pulsed muon source of the RIKEN-RAL muon
 facility \cite{riken-ral}, a tunable pulsed mid-infrared laser in the 6.8
 $\mu$m range \cite{famu-laser}, and a multi-pass cavity with mirrors of
 very high reflectance. The work on the optimization
 of the experimental set-up showed that in experiments with pulsed
 sources the time distribution of the laser energy in the
 multi-pass cavity is of primary importance. We present here a
 simplified model of the phenomena in this case that allows to
 qualitatively determine the optimal parameters of the cavity in
 dependence of the laser pulse parameters. Particular
 attention is paid to the optimal choice of the measurement time gate,
 i.e. the time interval in which the time distributions of the
 characteristic X-rays with and without laser are to be observed.

\section{Modelling the time profile of the electromagnetic fields
 in a multi-pass cavity}
\label{sec-1} 
\subsection{Preliminary remarks}
\label{sec-1.1}

 The probability $dp$ for the excitation of the ortho hyperfine
 ground state of $\mu^{-}p$ with a oscillating magnetic field of
 resonance frequency $\nu_0$
 in the time interval $dt$ such that $\nu_0\,dt\gg1$, under the condition
 that the width of the laser line is smaller that the Doppler broadening
 of the hyperfine transition line, can be put in the form:
 \begin{equation}
 dp=K |\vect{B}(\vect{r})|^2\,dt
 \end{equation}
 where $\vect{B}(r)$ is the amplitude of the magnetic field
 carried by the laser plane wave at the position $\vect{r}$ of the $\mu^{-}p$ atom, and
 the value of the dimension coefficient $K$ is expressed in terms
 of the masses and magnetic moments of the proton and the muon,
 the temperature and $\nu_0$ \cite{NIMB12}
 \begin{equation}
 K=\sqrt{\frac{(m_p+m_{\mu})c^2}{32\pi k T \nu_0^2}}
 \left(\frac{\mu_B}{\hbar}\left(\frac{m_e}{m_p}\mu_p+
 \frac{m_e}{m_{\mu}}\mu_{\mu}\right)\right)^2
 \end{equation}
 Accordingly, the probability $dP$ that a
 para-to-ortho transition will occur in the elemental
 volume $d^3\vect{r}$ (i.e. the expected number of laser-stimulated
 spin-flip events) is
 \begin{equation}
 dP=C(\vect{r})\,K
 |\vect{B}(\vect{r})|^2\,dt\,d^3\vect{r},
 \end{equation}
 where $C(\vect{r})$ is the number density of $\mu^{-}p$ atoms.
 At distances of the order of the wave length $\lambda_0=c/\nu_0$,
 the amplitude $\vect{B}(\vect{r})$ itself varies much faster than
 the atomic density $C(\vect{r})$,
 so we consider the probability $d\bar{P}$, obtained from
 $dP$ by averaging over a volume $\Delta V$ such that
 $\Delta V\gg\lambda_0^3$:
 \begin{equation}
 d\bar{P}/dt=
 \frac{1}{\Delta V}\int\limits_{\Delta V} \frac{dP}{dt}\,d^3\vect{r}=
 \frac{K}{\Delta V}\int\limits_{\Delta V}
 d^3\vect{r}\,C(\vect{r})\,|\vect{B}(\vect{r})|^2\approx
 K\,C(\vect{r}) \frac{1}{\Delta V}\int\limits_{\Delta V}
 d^3\vect{r}\,|\vect{B}(\vect{r})|^2
 = K\,C(\vect{r})\, \overline{|\vect{B}(\vect{r})|^2},
 \label{overline}
 \end{equation}
 where $\overline{|\vect{B}(\vect{r})|^2}$ is the amplitude of the
 magnetic field averaged at the wavelength scale.
 Denote by $N(t_1,t_2;V)$ the number of
 laser-stimulated spin-flip events that occur in the
 volume $V$ during the measurement time interval
 $[t_1,t_2]$:
 \begin{equation}
 N(t_1,t_2;V)=\int\limits_{t_1}^{t_2}dt \int\limits_{V}
 \frac{d\bar{P}}{dt}\,d^3\vect{r}=
 K \int\limits_{t_1}^{t_2}dt \int\limits_{V}
 C(\vect{r})\overline{|\vect{B}(\vect{r})|^2}\,d^3\vect{r},
 \end{equation}
 and by $N(t_1,t_2)$ -- the number of events in the whole cavity
 (of which, in the general case, only part of which is
 irradiated by the laser source).
 For uniform muonic atom density $C(\vect{r})=C_0$ this yields
 \begin{equation}
 N(t_1,t_2;V)=
 K\,C_0\int\limits_{t_1}^{t_2}dt \int\limits_{V}
 \overline{|\vect{B}(\vect{r})|^2}\,d^3\vect{r}.
 \end{equation}
 In a multi-pass cavity the oscillating magnetic field
 $\vect{B}(\vect{r},t)$ at $\vect{r}$ is the {\em incoherent sum}
 of the magnetic fields carried by the multiple reflected laser
 beams that irradiate the point $\vect{r}$:
 \begin{equation}
 \vect{B}(\vect{r},t)=\sum_n\vect{B}_n\exp i(\vect{k}_n.\vect{r}-\omega_0
 t+\delta_n)+c.c.,
 \end{equation}
 where $\vect{B}_n$, $\delta_n$ and $\vect{k}_n$ are the amplitude,
 phase and wave vector of the $n$-th plane wave, all of them with the same
 $\omega_0=2\pi\nu_0$ and $|\vect{k}_n|=k$.
 It can be shown for two summands that
 $\overline{|\vect{B}(\vect{r})|^2}=|\vect{B}_1|^2+|\vect{B}_2|^2+
 O((\theta k \, \Delta V)^{-1/3})$ (where $\theta$ is
 the angle between $\vect{k}_1$ and $\vect{k}_2$,
  $\vect{k}_1.\vect{k}_2=k^2\cos\theta$,
  and $\Delta V$ is the smoothing volume of Eq.(\ref{overline}));
 the latter is directly extended to any number
 of plane waves crossing at non-zero angles:
 \begin{equation}
 \overline{|\vect{B}(\vect{r})|^2}\approx\sum_n|\vect{B}_n|^2
 \label{incoh}
 \end{equation}

\subsection{A simple mathematical model of the multi-pass cavity}
\label{sec-1.2}

 We shall consider a simplified model of the multi-pass
 cavity aimed at qualitatively reproducing only the {\em time}
 dependence of $N(t_1,t_2;V)$. The evaluation of the {\em spatial} distribution of
 the spin-flip events requires specific simulation codes
 and the detailed knowledge of the $\mu^-p$-atom number density
 spatial distribution that are out of the scope of this talk;
 accordingly, we adopt the approximation of uniform number density
 of the muonic atoms, assume $V$ to be the {\em whole} cavity
 volume $\bar{V}$, and consider only the time distribution of
 $N(t_1,t_2)\equiv N(t_1,t_2,\bar{V})$.

\begin{figure}
 \centering
 \sidecaption
 \includegraphics[width=9cm,clip]{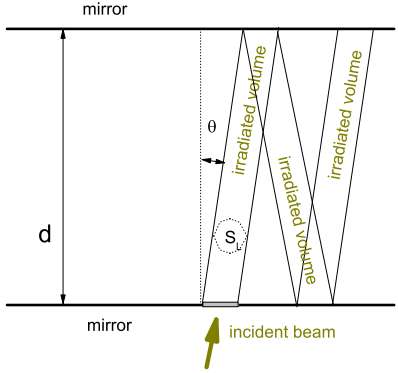}
 \caption{A schematic representation of the laser beam propagation
 in the optical multi-pass cavity. The distance between the two flat parallel
 mirrors is $d$; the beam incidence angle is $\vartheta$, and its cross section
 is $s_L$.}
 \label{f:the_cavity}       
 \end{figure}

 We assume that the cavity consists of two parallel mirrors at
 distance $d$. The laser beam with cross section $s_L$ is injected in
 the cavity at incidence angle $\alpha$ (see Fig.~\ref{f:the_cavity})
 and is multiply reflected by the mirrors of reflectance $R$. The details
 of the curved mirror edges that force the beam not to leave the
 cavity volume are not considered at all. The irradiated volume
 $V^*\in V$ may be thought to
 consists of cylindric segments with base surface $s_L$ and height
 $d/\cos\alpha\approx d$ for incidence angles $\alpha\ll\pi/2$;
 the time light travels through a segment is approximately
 $\tau_d=d/c$. We assume that the target gas is perfectly transparent
 for the laser light, so that $\overline{|\vect{B}(\vect{r})|^2}$
 does not vary along a segment (except for the zone of overlap
 of two segments -- an effect that can be accounted for at a later stage).
 At each reflection the value of
 $\overline{|\vect{B}(\vect{r})|^2}$ is suppressed by the factor
 $R$.

\begin{figure}[h]
 \centering
 \sidecaption
 \includegraphics[width=9cm,clip]{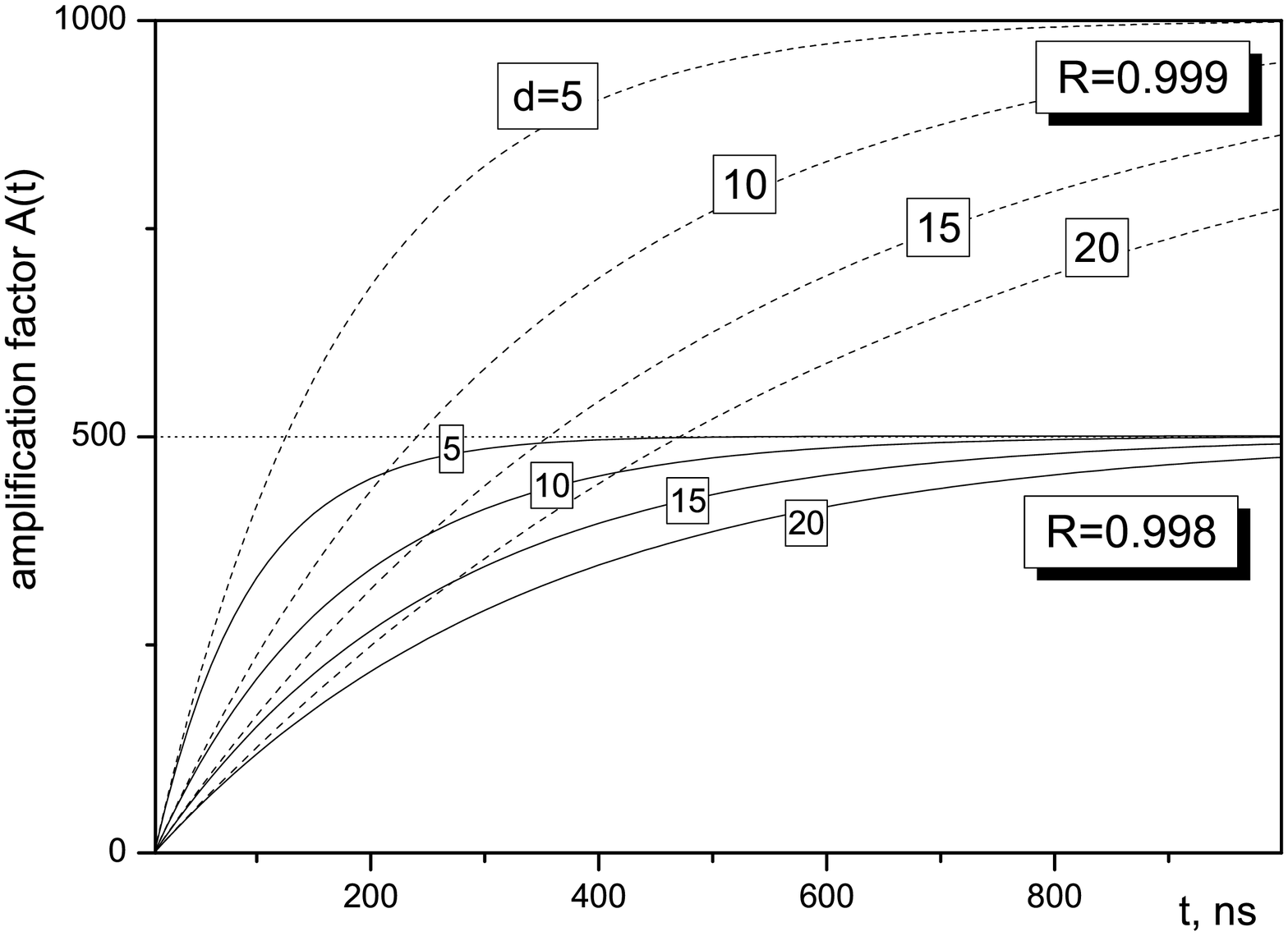}
 \caption{Amplification factor $A(t)$ of the number of laser-stimulated
 spin-flip events in a multi-pass cavity as function of the measurement
 time $t$, for laser pulse length $\tau_L=20$ ns, mirror
 reflectance $R=0.998$ and $R=0.999$, and distance between the mirrors
 $d=5(5)20$ cm.}
 \label{f:ampl0}       
 \end{figure}

 Let the laser pulse of duration $\tau_L$ enter the cavity at time
 $t_0=0$, and the the value of the averaged magnetic field at entrance
 be $\overline{|\vect{B}(\vect{r})|^2}=B_0^2$. Between $t=t_0$ and
 $t=\tau_L$ the irradiated volume $V$, as well as the number of events
 $N(t_0,t;V^*)$ increase
 monotonously: at each $\tau_d$ a new segment is irradiated. The
 value of $N(t_0,t;V)$ in this time interval is approximately
 \begin{equation}
 N(t_0,t)=
 \alpha
 \left(
 \frac{c}{1-R}\,t-d\,\frac{R(1-R^{tc/d})}{(1-R)^2}\right),\ \ 0\le
 t\le\tau_L,
 \label{t<}
 \end{equation}
 where $c$ is the speed of light and, for simplicity, we introduced
 the notation $\alpha=K\,C_0\,B_0^2\,s_L$.
 After the end of the laser pulse, for $t\ge\tau_L$, the
 irradiated spot propagates through the cavity but its volume
 $V$ remains unchanged, and after each
 reflection, i.e. at every $\tau_d$, the magnetic field in the
 segments is suppressed by the factor $R$, that leads to
 \begin{equation}
 N(\tau_L,t)=
 \alpha\,
 \frac{d}{(1-R)^2}\,
 \left(1-R^{c\tau_L/d}\right)
 \left(1-R^{c(t-\tau_L)/d}\right),
 \ \ t\ge\tau_L.
 \label{t>}
 \end{equation}
 From here $N(0,t)=N(0,\tau_L)+N(\tau_L,t)$.
 For large measurement times $t\gg\tau_d$
 \begin{equation}
 \lim_{t\to\infty}N(0,t)=
 \frac{\alpha}{1-R}\,c\tau_L+
 \frac{\alpha}{1-R}\,d\,
\left(1-R^{\tau_L/\tau_d}\right)
 \gtrsim
 \frac{\alpha}{1-R}\,c\tau_L.
 \label{large_t}
 \end{equation}
 Without a multi-pass cavity, the number of
 laser-stimulated spin-flip events at the same physical conditions
 would be $N_0=K\,C_0\,B_0^2\,s_l\,c',\tau_L=\alpha\,c\tau_L$. The ``amplification
 effect'' of the multi-pass cavity is therefore described with the factor
 \begin{equation}
 A(t)=N(0,t)/N_0=(1-R)^{-1}\times
 \left(1+\frac{\tau_d}{\tau_L}\left(1-R^{\tau_L/\tau_d}\right)
 \left(1-(1-R)^{-1}R^{(t-\tau_L)/\tau_d}\right)\right),\
 t\ge\tau_L,
 \label{ampl0}
 \end{equation}
 the whole dependence on $t$ being hidden in the rightmost factor.
 Fig.~\ref{f:ampl0} is a plot of $A(t)$ for the typical values
 of the parameters $\tau_L=20$ ns, $R=0.998$ and $R=0.999$, and the set of
 values $d=5(5)202$ cm.

 \section{Optimization of the measurement time gate}
 \label{sec-3}

\begin{figure}[h]
 \centering
 \sidecaption
 \includegraphics[width=9cm,clip]{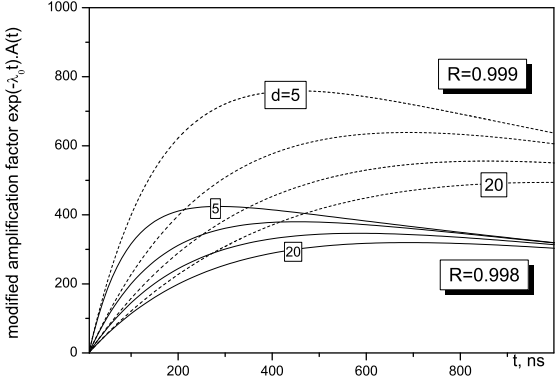}
 \caption{Modified amplification factor $\exp(-\lambda_0t)A(t)$
 of the number of laser-stimulated
 spin-flip events in a multi-pass cavity as function of the measurement
 time $t$, for laser pulse length $\tau_L=20$ ns, mirror
 reflectance $R=0.998$ and $R=0.999$, and distance between the mirrors
 $d=5(5)20$ cm.; $\lambda_0=0.45\mu{\rm s}^{-1}$ is the muon decay rate.}
 \label{f:ampl1}       
 \end{figure}

 Eq.~(\ref{large_t}) gives the upper limit of the amplification
 factor: $A(t)\simeq(1-R)^{-1}$; Fig.~\ref{f:ampl0} shows that the time
 needed to reach the maximal amplification grows fast with the
 distance $d$ between mirrors and with the mirror reflectivity
 $R$.
 Knowing
 the explicit dependence of the number of spin-flip events on the
 measurement time $t$ may therefore help select the
 optimal measurement time gate $[0,t]$ for which the number of
 laser-stimulated spin-flip events is maximal.
 In such an optimization problem one should, of course, take into
 account the finite muon decay rate $\lambda_0$ as well. As shown of
 Fig.~\ref{f:ampl1}, the muon decay partly suppresses the
 gain of laser-stimulated spin-flip events due to the multi-pass cavity.

 To optimize the efficiency of the FAMU experiment, however,
 one should maximize the  {\em signal-to-noise
 ratio $\Delta/\sigma$} rather than just the number
 of  laser-stimulated spin-flip events.
 In the experimental method of FAMU,
 the ``noise'' $\sigma$ (also referred to as ``background'')
 mainly consists of characteristic X-rays
 emitted during the relaxation of the $\mu^-O$ atoms formed
 through muon transfer from {\em thermalized} muonic hydrogen
 atoms (and therefore independent of the laser light), while
 the ``signal'' $\Delta$ is the variation of the the background
 distribution due to X-rays from $\mu^-O$ atoms formed by muon
 transfer from {\em epithermal} $\mu^-H$  \cite{hfi2k,NIMB12}.
 Obviously, the latter is
 proportional to the number of laser-stimulated events,
 while the ``noise'' depends
 only on the number of muonic atoms in target:
 \begin{equation}
 \Delta\sim \exp(-\lambda_0t)A(t),\
 \sigma^2\sim\int\limits_0^t\exp(-\lambda_0\,t')\,dt'=
 (1-\exp(-\lambda_0t))/\lambda_0,
 \end{equation}
 so that the time dependence $S(t)$ of the signal-to-noise ratio is:
 \begin{equation}
 S(t)=\frac{\Delta}{\sigma}\sim\frac{\exp(-\lambda_0t)}
 {\sqrt{1-\exp(-\lambda_0t))}}\,A(t).
 \label{s/n}
 \end{equation}
 The noise $\sigma$ grows monotonously with the duration $t$
 of the measurement time gate while
 the signal $\Delta$ reaches its maximum at some finite moment
 (see Fig.~\ref{ampl1}). We can
 therefore expect $S(t)$ to have pronounced maxima
 for any value of $d$ and $R$, that will determine
 the optimal time gate for measurement of the hyperfine
 splitting of muonic hydrogen in the FAMU approach.

 \begin{figure}[h]
 \centering
 \sidecaption
 \includegraphics[width=9cm,clip]{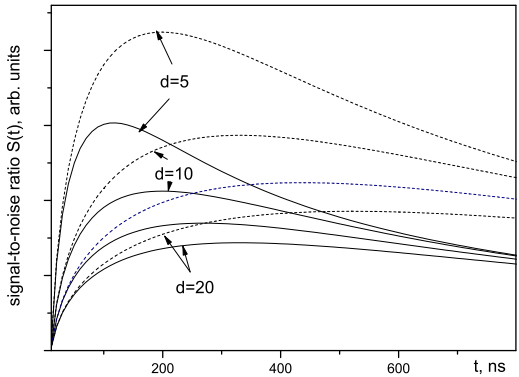}
 \caption{Dependence of the signal-to-noise ratio
 $S(t)=\Delta/\sigma$ (in arbitrary units) on the measurement time
 gate $t$, for a set of inter-mirror distances $d=5(5)10$ cm, mirror
 reflectivity $R=0.998$ (solid lines) or $R-0.999$ (dashed lines),
 and laser pulse length $\tau_L$=20 ns. }
 \label{f:stn}       
 \end{figure}

 Fig.~\ref{f:stn} confirms these qualitative predictions.
 We also note that:
 \begin{itemize}
 \item The maximal achievable signal-to-noise ratio is increased
 as the distance $d$ between the mirrors decreases;
 \item The optimal time gate length $t$ grows with inter-mirror
 distances $d$;.
 \end{itemize}

 \section{Conclusions}
 \label{sec-4}

 The simplified model of the optical multi-pass cavity considered
 here completely neglects the effects of the spatial distribution of the muonic
 hydrogen atoms and of the laser
 radiation throughout the cavity,
 as well as some other details of the undergoing physical processes,
 and cannot be directly applied in simulations of the FAMU
 experiment. Still, it reveals important peculiarities
 related to the use of multi-pass cavities in experiments with
 pulsed laser and muon sources, and can be helpful in the
 preliminary design of the experimental set-up and the
 approximate choice of the optimal measurement time gate and
 inter-mirror distance.

 \section{Acknowledgments}

 The authors acknowledge the support of Grant 08-17 of the
 Bulgarian Science Fund.

%
%

\end{document}